\documentclass{emulateapj-rtx4}
\usepackage{apjfonts}
\usepackage[usenames]{color}

\newcommand{\gz}{$g\!-\!z$}

\newcommand{\Modot}{M_{\odot}}

\newcommand{\cM}{{\cal{M}}}
\newcommand{\ML}{${\cal{M}}/L$}
\newcommand{\mybibitem}[3]{\bibitem[{#1}({#2})]{#3}}
\newcommand{\mybibthree}[4]{\bibitem[{#2}({#3}){#1}]{#4}}
\newcommand{\picplace}[1]{\vbox{\hrule\@height 0.4pt\@width\hsize
\hbox to\hsize{\vrule\@width 0.4pt\@height#1\hfil
\vrule\@width 0.4pt\@height#1}\hrule\@height 0.4pt\@width\hsize}}

\slugcomment{Accepted for Publication by ApJ}

\shorttitle{The Color-Magnitude Relation among GCs in Massive Galaxies}
\shortauthors{P. Goudfrooij \& J. M. D. Kruijssen}

\begin{document}


\title{Color-Magnitude Relations within Globular Cluster Systems of Giant
  Elliptical Galaxies: \\ The Effects of Globular Cluster Mass Loss and the
  Stellar Initial Mass Function}     



\definecolor{MyBlue}{rgb}{0.3,0.3,1.0}

\author{Paul Goudfrooij$^1$ and J. M. Diederik Kruijssen$^2$ \vspace*{0.3mm}}
\affil{$^1$ Space Telescope Science Institute, 3700 San Martin
  Drive, Baltimore, MD 21218, USA; {\color{MyBlue}goudfroo@stsci.edu} \\ 
  $^2$ Max-Planck Institut f\"ur Astrophysik, Karl-Schwarzschild-Str.\ 1,
  D-85741 Garching, Germany; {\color{MyBlue}kruijssen@mpa-garching.mpg.de}} 






\begin{abstract}
Several recent studies have provided evidence for a ``bottom-heavy'' stellar
initial mass function (IMF) in massive elliptical galaxies. Here we
investigate the influence of the IMF shape on the recently discovered
color-magnitude relation (CMR) among globular clusters (GCs) in such
galaxies. 
To this end we use calculations of GC mass loss due to stellar and dynamical
evolution to evaluate 
\emph{(i)} the shapes of stellar mass functions in GCs after 12 Gyr of
evolution as a function of current GC mass along with their effects on
integrated-light colors and mass-to-light ratios, and  
\emph{(ii)} their impact on the effects of GC self-enrichment using the
2009 ``reference'' model of Bailin \& Harris. 
As to the class of metal-poor GCs, we find the observed shape of
the CMR (often referred to as the ``blue tilt'') to be very well reproduced by
Bailin \& Harris' reference self-enrichment model once 12 Gyr of GC mass loss
is taken into account. The influence of the IMF on this result is found to be
insignificant. 
However, we find that the observed CMR among the class of \emph{metal-rich} 
GCs (the ``red tilt'') can only be adequately reproduced if the IMF was
bottom-heavy ($-3.0 \la \alpha \la -2.3$ in $dN/d\cM \propto \cM^{\alpha}$),
which causes the stellar mass function at subsolar masses to depend 
relatively strongly on GC mass. This constitutes additional evidence that the
metal-rich stellar populations in giant elliptical galaxies were formed with a
bottom-heavy IMF. 

\end{abstract}






\section{Introduction}              \label{s:intro}

Globular clusters (GCs) are thought to be formed in bursts of star formation
within molecular protocluster clouds, producing gravitationally bound clusters
of 10$^4$\,--\,10$^7$ stars which represent the best known approximations of a
``simple stellar population'' (hereafter SSP), i.e., a coeval population of
stars with a single metallicity \citep[e.g.,][and references
therein]{brostr06}. 
Extensive photometric observations of extragalactic GC systems have now 
yielded a substantial database of GC colors and
magnitudes to investigate overall trends among subpopulations of GCs and
between properties of GC subpopulations and those of their host
galaxies. The advent of the \emph{Advanced Camera for Surveys (ACS)} aboard
the \emph{Hubble Space Telescope (HST)} was pivotal in this area due to
its unique high sensitivity and spatial resolution over a field of view of
$\simeq$\,12 arcmin$^2$
\citep[e.g.,][]{cote+04,goud+04,goud+07,harr+06,harr+09,jord+07a,peng+06a,peng+11}.  

One main result of these studies was that GCs in luminous galaxies typically
follow a bimodal color distribution.  Multi-object spectroscopy
studies showed that both ``blue'' and ``red'' GC subpopulations are
nearly universally old, with ages $\ga$\,8 Gyr
\citep[e.g.,][]{cohe+03,puzi+05}. This implies that the color
bimodality is mainly due to differences in metallicity \citep[see
  also][]{brod+12}. Typical
metallicity values for metal-poor (blue) GCs in giant elliptical
galaxies are [Z/H] = $-1.3 \pm 0.3$, while metal-rich (red) GCs
average [Z/H] = $-0.2 \pm 0.3$ \citep{harr+06,peng+06a}. 

A more recently discovered feature among GC systems was a color-magnitude 
relation among GCs in the metal-poor subpopulation, in 		
the sense that their colors become progressively redder at higher
luminosities \citep{harr+06,harr+10,stra+06,spit+06,mies+06,mies+10,forb+10}.

This relation, colloquially called the ``blue tilt'', is typically interpreted
as a mass-metallicity relation (hereafter MMR), although the various analyses
of the effect in individual galaxies have not yet converged to an agreement on
the significance of the tilt and the mass range in which the
correlation is significant (see, e.g., \citealp{kund08} and \citealp{wate+09}
versus \citealp{harr+06}, \citealp{mies+06}, and \citealp{peng+09}). 
Recently, \citet{mies+10} performed a detailed investigation of the
color-magnitude relation in the largest sample of GCs to date by far
($\approx$\,16,900 GCs, see Section~\ref{s:sample}), using $g$ and $z$
filters. Three main conclusions of \citet{mies+10} regarding the blue tilt
were that  {\it (i)\/} the blue tilt is firmly present for GC masses $\cM_{\rm
  GC} \ga 2 \times 10^5 \; M_{\odot}$ while it becomes more pronounced at $\cM_{\rm
  GC} \ga (2-3) \times 10^6 \; M_{\odot}$;  
{\it (ii)\/} the tilt is more pronounced for GCs associated with high-mass galaxies 
($\cM_{\rm gal} \ga 5 \times 10^{10} \; M_{\odot}$) than for GCs in lower-mass 
galaxies; and 
{\it (iii)\/} the tilt is more pronounced for GCs at smaller galactocentric
distances. 
Notably, \citet{mies+10} also found evidence for the presence of a ``red
tilt'', i.e., a color-magnitude relation among GCs in the metal-rich
subpopulation. This red tilt is seen among GCs in the full magnitude range
sampled by \citet{mies+10}, albeit only within their ``high-mass'' subset of galaxies. 

The vast majority of studies of the MMR among GCs explain it as being due to
GC self-enrichment by ejecta of supernovae of type II
\citep[SN\,II; e.g.,][]{strsmi08,baihar09}. However, several features of the
blue tilt mentioned above seem to indicate that GC self-enrichment alone
cannot fully explain the observed MMR. For example, most theoretical work
on GC self-enrichment in heavy (Fe-peak) elements indicates that it is
only effective at masses $\cM_{\rm GC} \ga 2 \times 10^6 \; M_{\odot}$ 
\citep[e.g.,][]{morlak89,recdan05,fell+06,baihar09} 
whereas the blue tilt in elliptical galaxies persists down to GC masses
that are an order of magnitude lower\footnote{We note that star-to-star abundance
  variations among {\it light} elements (up to $^{13}$Al) are commonly found
  within Galactic GCs, and there are indications that the
  amplitude of such variations scale with GC mass
  \citep[e.g.,][]{carr+10a,grat+12}. However, such variations only cause a 
  small blueing of integrated \gz\ colors \citep[$0.00 \ga \Delta g-z \ga
  -0.03,$][]{goukru13}, which cannot cause the blue or red tilts.}.  
Furthermore, one would not expect to see significant environmental dependences
of the slope of the MMR (such as the tilt being more pronounced for
higher-mass galaxies or for smaller galactocentric distances) if GC
self-enrichment were the only relevant cause. Instead, the presence of such
environmental variations seem to indicate an effect 
(or effects) whose strength is driven by the local tidal field, such as
star cluster disruption.   

Recent investigations of the distribution of Fe-peak element
abundances within Galactic GCs are starting to yield clues in 
this regard. Statistically significant internal variations in [Fe/H]  
have so far been detected within a few (metal-poor) GCs: 
$\omega$\,Cen \citep[e.g.,][]{nordac95,johpil10}, 
M\,54 \citep{sarlay95,bell+08,carr+10b}, and M\,22 \citep{marino+11}. These GCs
have current masses of $2.3 \times 10^6 \; M_{\odot}$, $2.0 \times 10^6 \;
M_{\odot}$, and $2.6 \times 10^5 \; M_{\odot}$ respectively \citep{mclvdm05}.  
Note that the mass of M\,22 is comparable to the onset of the blue tilt found by
\citet{mies+10}, lending credence to the potential relevance of GC mass loss
in this context. 

Meanwhile, recent studies of the central regions of nearby massive
elliptical galaxies revealed very strong gravity-sensitive features in
their near-IR spectra \citep{vancon10,convan12b}, indicating a stellar
initial mass function (IMF) that was steeper at subsolar masses
($\alpha \simeq -3$ in $dN/d\cM \propto \cM^{\alpha}$) than the
canonical \citet{krou01} or \citet{chab03} IMFs, which have $\alpha
\simeq -1.7$ in the same mass range. This was subsequently found to be
consistent with studies of stellar \ML\ ratios of massive early-type
galaxies derived from gravitational lensing \citep[e.g.,][]{sonn+12}
and stellar dynamics \citep{capp+12,tort+13}. 
In the same vein, \citet{goukru13} reported a systematic
offset between optical colors of giant elliptical galaxies and the
mean colors of their massive metal-rich GCs, in the sense that those
GCs were bluer than their parent galaxies by 0.12\,--\,0.20
mag at a given galactocentric distance (a 7\,--\,22 $\sigma$
effect). However, measurements of Lick-system spectral indices that are
sensitive to age and metallicity (but not to IMF variations) for such 
galaxies and their metal-rich GCs {\it were\/} found to be consistent with one
another at a given galactocentric distance, within small uncertainties. 
\citet[][hereafter Paper I]{goukru13} showed that this paradox can be explained
if the stellar IMF was steep at subsolar masses: $\sim$\,12 Gyr of 
dynamical evolution would then cause the average massive GC to become bluer by an amount
consistent with the observed color difference between the metal-rich GCs and
their parent galaxies.  

To investigate these recent indications of the relevance of the IMF shape
  and dynamical evolution with regards to the colors of GCs and features of the
  observed MMR among GCs in nearby galaxies, we conduct a study of the 
effects of mass loss of GCs born with different IMF shapes to the MMR among
GCs in giant elliptical galaxies and compare those effects to the observed MMR
and the effect of self-enrichment.  

\section{The Mass-Metallicity Relation among Globular Clusters} \label{s:sample}

In comparing results of our calculations with observed features of the GC MMR
in giant elliptical galaxies, we adopt the parametrizations of the
observed color-magnitude relation and the MMR among metal-poor and
metal-rich GCs by \citet[][see their Table 1]{mies+10}. This choice
was made because their  MMR study gathered the largest photometric
sample of GCs to date by far ($\approx$\,16,900 GCs), namely by
combining all GCs detected in  the ACS Fornax and Virgo Surveys of
early-type galaxies. These {\it HST\/} surveys were carried out
using a homogeneous measurement technique in which photometric
aperture-size corrections were determined from convolutions of the
best-fit GC profiles with local point-spread functions on the image
(see \citealp{jord+09} for details).  

Briefly, GC masses and metallicities in \citet{mies+10} were derived from 
HST/ACS photometry in the F475W (hereafter $g$) and F850LP ($z$)
filters, corrected for Galactic foreground extinction. Metallicities (in terms
of [Fe/H] on the \citet{zinwes84} scale) were derived using the broken linear 
transformation of \citet{peng+06a}, while GC masses were derived from $z$-band
absolute magnitudes, using $\cM/L_z = 1.5 \; M_{\odot}/L_{z,{\odot}}$,
which is correct to within 5\% in the range $-2.2 \la 
\mbox{[Fe/H]} \la +0.2$ for an SSP \citep{jord+07b}. 

We refer to \citet{mies+10} for further details regarding the
determination of color-magnitude relations and MMRs among their various
subsamples of GCs.

\section{Dynamical Evolution Effects on the Color-Magnitude Relation among
  Globular Clusters} \label{s:evol} 

\subsection{Dynamical Evolution and Stellar Mass Functions} \label{s:MFevol}

It is well known that mass loss of star clusters by dynamical evolution affects
the shape of their stellar MF, because the escape probability of stars from
their parent cluster increases with decreasing stellar mass
\citep[e.g.,][]{heno69,baumak03,krui09}. 
This effect is widely used to explain observed stellar MFs in
ancient Galactic GCs, some of which are flatter than canonical IMFs
\citep[e.g.,][]{dema+07,paus+10}.  
Such flat MFs in GCs can explain their low observed
mass-to-light ratios when compared with predictions of SSP
models that use canonical IMFs \citep[e.g.,][]{krupor09}. 

\begin{figure*}[tbhp]
\centerline{
\includegraphics[width=14.cm]{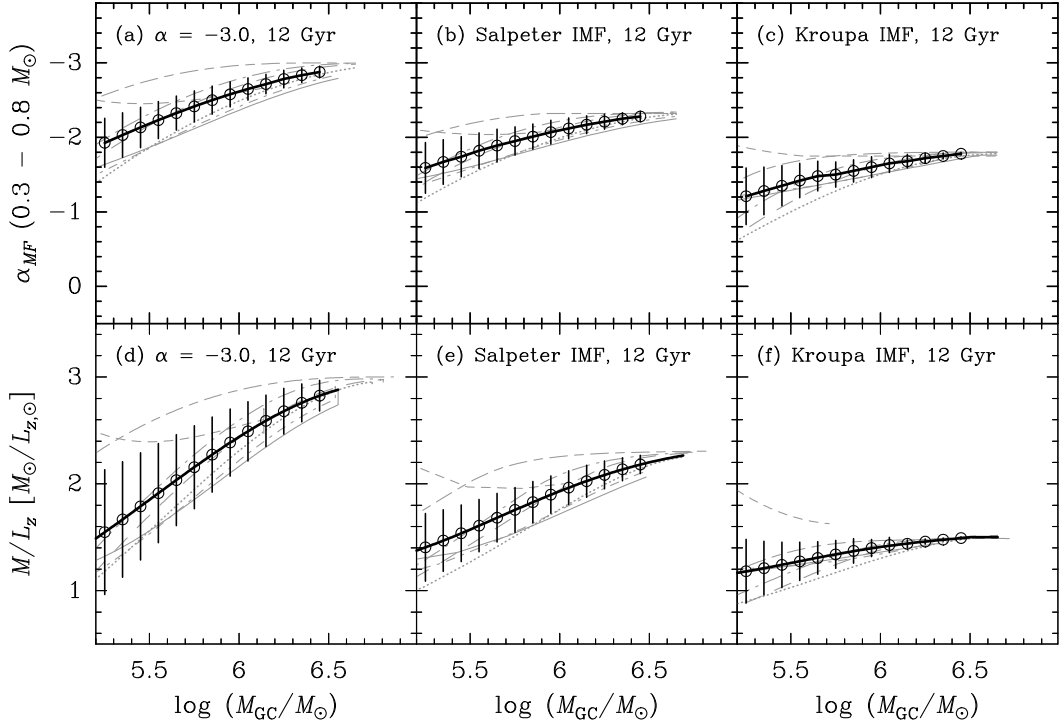}
}
\caption{{\it Panel (a)}: MF slope $\alpha_{\it MF}$ in the range $0.3
  < \cM_{\star}/\Modot < 0.8$ versus GC mass at an age of 12 Gyr 
  for all models of K09 that use $\alpha_{\it IMF} = -3.0$ and produce
  surviving GCs with $\cM_{\rm GC} \geq 2.5 \times 10^5 \: \Modot$ at an age
  of 12 Gyr. Different line styles in grey indicate K09 models with the
  following combinations of ($t_0$, $W_0$):\ 
  solid lines (0.3, 7); short dashes (0.6, 5); dots (0.6, 7); 
  short dash-dot-dashes (1.0, 5); long dashes (1.0, 7); long dash-dot-dashes
  (3.0, 5); and long-short-long dashes (3.0, 7).  
  The thick solid curve in black represents the average of all curves
  mentioned above; 1$\sigma$ uncertainties 
  are overplotted using bins of log $\cM_{\rm GC}$ = 0.1. 
  {\it Panel (b)}: Similar to panel (a), but now for a \citet{salp55} IMF
  ($\alpha_{\it IMF} = -2.35$). 
  {\it Panel (c)}: Similar to panel (a), but now for a \citet{krou01} IMF.
  {\it Panel (d)}: $\cM/L_z$ ratio as function of GC mass at an age of
  12 Gyr. Grey lines of various styles represent the same K09 
  models for $\alpha_{\it IMF} = -3.0$ as those shown in panel (a)
  with the same line styles.   
  Open circles and black lines represent the ``average'' models of K09
  that were shown with black lines in panel (a). 
  {\it Panel (e)}: Similar to panel (d), but now for a \citet{salp55} IMF. 
  {\it Panel (f)}: Similar to panel (d), but now for a \citet{krou01} IMF.
\label{f:MF_and_M_Lz_vs_logM}
}
\end{figure*}

To evaluate the influence of the shape of the IMF on the present-day
color-magnitude relation among GCs, we follow Paper I and use the GC evolution
model of \citet[][hereafter K09]{krui09} which incorporates the effects of
stellar evolution, dynamical evolution in a tidal field, mass segregation, and stellar
remnant retention. For the purposes of the current study, we select K09 models 
that {\it (i)\/} feature ``default'' kick velocities of stellar remnants
(white dwarfs, neutron stars, and black holes), and {\it (ii)\/} produce GCs
with masses $\cM_{\rm GC} \geq 2 \times 10^5 \: \Modot$ at an age of 12 Gyr,
i.e., GCs such as those selected by \citet{mies+10} to derive their MMRs. 
We further consider \citet{king66} profiles with $W_0$ values of 5 and 7 and
environmentally dependent cluster dissolution time scales $t_0$ of 0.3, 0.6,
1.0, and 3.0 Myr\footnote{$t_0$ is defined by $t_{\rm dis} = t_0 \, (\cM_{\rm
    GC}/\Modot)^{\gamma}$ where $t_{\rm dis}$ is the cluster disruption time and
  $\gamma$ sets the mass dependence of cluster disruption. See \citet{lame+05}
  and K09 for details.}. For reference, $t_0 = 1.3$ Myr yields a good fit to the
globular cluster mass function of all surviving Galactic GCs \citep{krupor09}. 
However, rather than using a single $t_0$ value for a given
galaxy, we consider that GC systems in galaxies are expected to exhibit a
\emph{range} of $t_0$ values. 
For instance, the rate of mass loss from GCs by two-body relaxation,  
$\mu_{\rm rel}$, effectively scales with the mean GC half-mass density
$\rho_{\rm h}$ as $\mu_{\rm rel} \propto \rho_{\rm h}^{0.5}$ for tidally
limited GCs (e.g., \citealt{mclfal08,giel+11,goud12}). Among GCs for which 
two-body relaxation has been the dominant mass loss mechanism, 
low-$\rho_{\rm h}$ GCs thus feature larger $t_0$ values than
high-$\rho_{\rm h}$ GCs at a given $\cM_{\rm GC}$. Similarly, GCs on eccentric 
orbits with large perigalactic distances $R_{\rm peri}$ should have larger $t_0$
values than otherwise (initially) similar GCs with smaller values of $R_{\rm
  peri}$.  
Finally, disruption rates of GCs in dense environments such as central
regions of gas-rich galaxy mergers at high redshift were likely higher
than in current quiescent galaxies \citep{fall+09,krui+12} due to the enhanced
strength and rate of tidal perturbations.  

In Paper I, we explored the influence of the IMF on the
present-day MF at subsolar masses for the star cluster families mentioned
above. For an ensemble of such GCs, our calculations showed that the overall
average \emph{present-day} MF slope in the stellar mass range 0.3\,--\,0.8
$M_{\odot}$ (hereafter designated $\alpha_{\it MF}$) is consistent with the
Kroupa MF for any IMF slope $-3.0 \la \alpha_{\it IMF} \la -2.0$ (see Figure 7
of Paper I). However, the {\it mass dependence\/} of $\alpha_{\it MF}$
(averaged over all K09 models that produce GCs of a given mass), is found to be
significantly stronger for steeper IMFs. This is illustrated in
Figures~\ref{f:MF_and_M_Lz_vs_logM}a\,--\,\ref{f:MF_and_M_Lz_vs_logM}c,
which show $\alpha_{\it MF}$ versus $\cM_{\rm GC}$ for all K09 models that
produce GCs with $\cM_{\rm GC} \geq 2 \times 10^5 \; M_{\odot}$ at an age of
12 Gyr, as well as the average model, for three IMFs: 
\emph{(i)} A ``bottom-heavy'' IMF with $\alpha_{\it IMF} = -3.0$ for $\cM/M_{\odot} <
0.8$ and $\alpha_{\it IMF} = -2.35$ for $0.8 \leq \cM/M_{\odot} \leq 100$ 
(this IMF is hereafter simply referred to as $\alpha_{\it IMF} = -3.0$), 
\emph{(ii)} the \citet{salp55} IMF (i.e., $\alpha_{\it IMF} = -2.35$ over the full
stellar mass range), and 
\emph{(iii)} the \citet{krou01} IMF. 
As described in detail in Paper I, the stronger cluster mass dependence of
$\alpha_{\it MF}$ among ancient GCs born with a steeper IMF is mainly caused
by the weaker effect of retained massive stellar remnants on the escape rate
of massive stars in GCs with steeper IMFs relative to that in GCs with flatter
IMFs.   

The effect of 12 Gyr of dynamical evolution on the $\cM/L_z$ ratio as a
function of $\cM_{\rm GC}$ is similar to that on $\alpha_{\it MF}$. This is
shown in
Figures~\ref{f:MF_and_M_Lz_vs_logM}d\,--\,\ref{f:MF_and_M_Lz_vs_logM}f. As 
expected, steeper IMFs feature higher $\cM/L_z$ ratios for the most massive
GCs. However, steeper IMFs also feature a stronger cluster mass dependence of
$\cM/L_z$, and a larger spread of $\cM/L_z$ ratios among different GC families
at a given current mass. This result renders it difficult to constrain the IMF
slope for individual ancient GCs through direct measurements of $\alpha_{\it
  MF}$ or $\cM/L$, especially for GCs with $\log \cM_{\rm GC} \la
5.5$. However, the large spread of observed MF slopes and $\cM/L$
ratios among Galactic GCs \citep{dema+07,paus+10,zari+13} generally appears more
consistent with a bottom-heavy IMF than with a Kroupa or Chabrier one. 
In the remainder of this paper, the cluster mass -- $\cM/L_z$ relations shown
by the thick solid lines in
Figs.~\ref{f:MF_and_M_Lz_vs_logM}d\,--\,\ref{f:MF_and_M_Lz_vs_logM}f will be
taken into account when comparing model predictions to the observed MMR from
\citet{mies+10}, since the latter study used $\cM/L_z = 1.5 \;
M_{\odot}/L_{z,{\odot}}$ independent of cluster mass.  

\subsection{Dynamical Evolution and $g-z$ Colors} \label{s:gzevol}
\begin{figure}[tbhp]
\centerline{
\includegraphics[width=5.7cm]{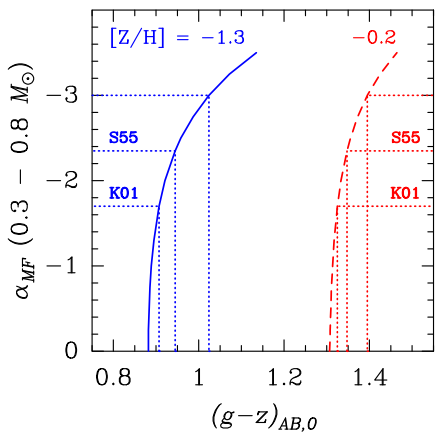}
}
\caption{Integrated \gz\ color vs.\ $\alpha_{\it MF}$ using 
  \citet{mari+08}
  isochrones for an age of 12 Gyr and the two [Z/H] values
  $-$1.3 (solid line) and $-$0.2 (dashed line). 
  For convenience, dotted lines indicate values for \gz\ and
  $\alpha$ for the Kroupa IMF (K01), the Salpeter IMF (S55) and an IMF
  with $\alpha = -3.0$.  
\label{f:color_vs_alpha}
}
\end{figure}

The impact of $\alpha_{\it MF}$ on the integrated \gz\ color of ancient (12 Gyr
old) GCs is shown in Figure~\ref{f:color_vs_alpha} for [Z/H] = $-$1.3 
and [Z/H] = $-$0.2, i.e., the two typical metallicity values for the metal-poor
and metal-rich GC subpopulations in giant elliptical galaxies, respectively. 
These curves were determined from Padova isochrones \citep{mari+08}. After
rebinning the isochrone tables to a uniform bin size in stellar mass
$\cM_{\star}$ using linear interpolation, weighted luminosities of the
population were derived for the relevant passbands by normalizing individual
stellar luminosities $L_{\star}$ as follows: 
\begin{eqnarray}
L_{\star,W} & = & 
  L_{\star} \left({\cM_{\star} \over \cM_{\rm max}}\right)^{-2.35} 
 \qquad \qquad \,{\rm if~} \cM_{\star} \geq \cM_{1} \\
 & = & 
  L_{\star}  \left({\cM_1 \over \cM_{\rm max}}\right)^{-2.35}  
  \left({\cM_{\star} \over \cM_1}\right)^{\alpha} \; 
  {\rm if~} 0.1 \; M_{\odot} < \cM_{\star} \leq \cM_1, \nonumber 
\end{eqnarray}
where $\cM_{\rm max}$ is the maximum stellar mass reached in the isochrone
table, $\cM_1 = 0.8 \; \Modot$, and $\alpha$ is the MF slope in
the mass range $0.1\;\Modot < \cM_{\star} < \cM_1$.  

\begin{figure}[tbhp]
\centerline{
\includegraphics[width=8.3cm]{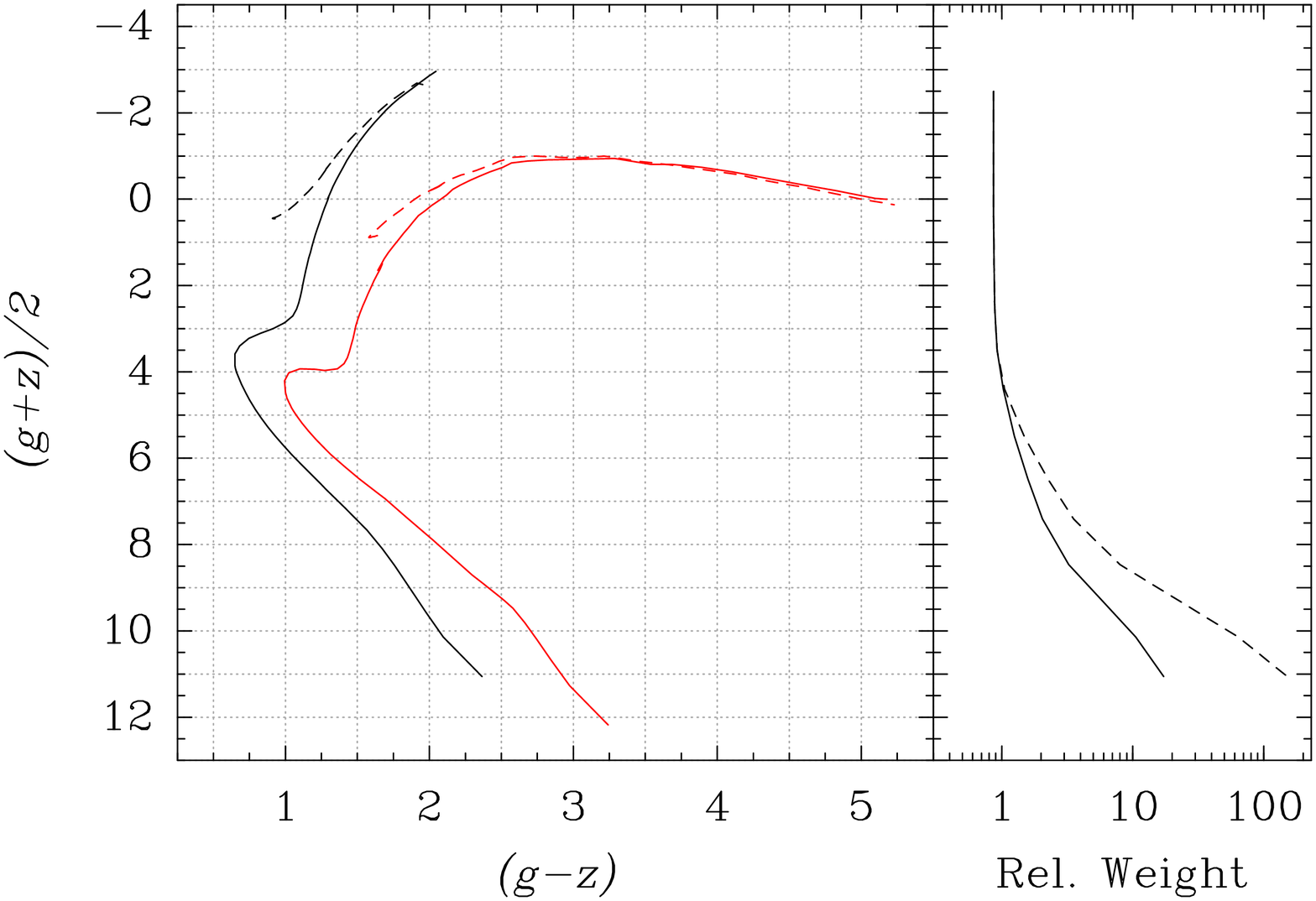}
}
\caption{\emph{Left panel}: Comparison of \citet{mari+08} isochrones of
  age 12 Gyr and metallicities [Z/H] = $-$1.3 (black lines) and $-$0.2 (red lines) in a
  $(g+z)/2$ vs.\ $g-z$ color-magnitude diagram. The AGB phase is indicated with
  dashed lines. 
  \emph{Right panel}: Illustration of the relative weights of stellar
  luminosities implied by two mass functions: the Kroupa IMF (solid line) and
  $\alpha_{\it IMF} = -3.0$ (dashed line). 
  See discussion in Section~\ref{s:gzevol}.   
\label{f:isocomparison}
}
\end{figure}

Note that the effect of the IMF shape on the \gz\ color is larger for
metal-poor populations than for metal-rich ones. 
This difference can be understood by considering the shape of
the isochrones in color-magnitude diagrams (CMDs) at different
metallicities. This is illustrated in Figure~\ref{f:isocomparison} which shows
\citet{mari+08} isochrones for [Z/H] = $-$1.3 and [Z/H] = $-$0.2 (at an
age of 12 Gyr) in a ($g+z$)/2 vs.\ \gz\ CMD\footnote{($g+z$)/2 was chosen as
  the ordinate for the CMD in order to show the luminosity exactly in between
  the $g$ and $z$ passbands.}. 
Note that metal-rich populations feature a strong curvature of the red giant
branch (RGB) in these passbands, rendering \gz\ colors of upper RGB stars that
are significantly redder than those of lower main sequence (MS)
stars. Conversely, the curvature of the RGB in the CMD for metal-poor
populations is much weaker, and upper RGB stars are {\it bluer\/} than lower
MS stars for metal-poor populations. As the overall range of ($g+z$)/2
magnitudes spanned by the isochrones is similar for metal-poor and metal-rich
populations, the impact of the MF slope at subsolar masses on the integrated
colors is larger for metal-poor populations than for metal-rich ones.  

\begin{figure*}[tbhp]
\centerline{
\includegraphics[width=13.5cm]{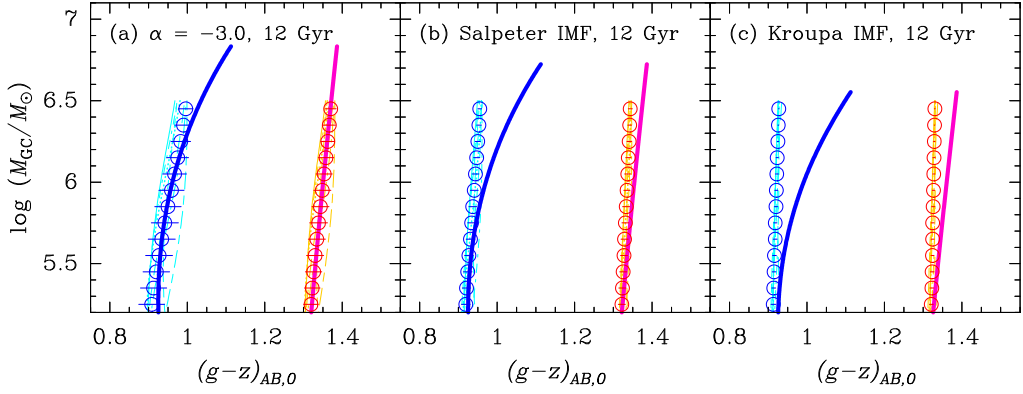}
}
\caption{GC mass at an age of 12 Gyr versus \gz. {\it Panel (a)}: Thin lines of
  various styles represent the same K09 models for $\alpha_{\it IMF} =
  -3.0$ as those shown in Fig.\ \ref{f:MF_and_M_Lz_vs_logM}a with the same line
  styles.  
  Open circles represent the ``average'' models of K09 which were shown with
  black solid lines in Fig.~\ref{f:MF_and_M_Lz_vs_logM}a.   
  Blue symbols or cyan lines represent models for metal-poor GCs with [Z/H] =
  $-$1.3 while red symbols or orange lines represent models for metal-rich GCs
  with [Z/H] = $-$0.2.  
  For comparison, the thick lines represent the observed color-mass relations
  among metal-poor (blue line) and metal-rich (magenta line) GCs in giant
  elliptical galaxies by \citet[][see bottom left panel of their
  Fig.~5]{mies+10}, after accounting for the $\cM/L_z$ -- $\cM_{\rm GC}$
  relations shown by the solid black lines in
  Fig.~\ref{f:MF_and_M_Lz_vs_logM}d. 
  {\it Panel (b)}: Similar to panel (a), but now for a \citet{salp55} IMF
  ($\alpha_{\it IMF} = -2.35$). 
  {\it Panel (c)}: Similar to panel (a), but now for a \citet{krou01} IMF.
\label{f:Mcurr_vs_color}
}
\end{figure*}

To transform the GC mass-- and IMF-dependent effect of GC disruption
on $\alpha_{\it MF}$ at an age of 12 Gyr seen in
Figs.~\ref{f:MF_and_M_Lz_vs_logM}a\,--\,\ref{f:MF_and_M_Lz_vs_logM}c 
to observable properties, we apply the $\alpha_{\it MF}$ vs.\ \gz\ relations shown in
Figure~\ref{f:color_vs_alpha} to the K09 models for the three IMFs
shown in Figure~\ref{f:MF_and_M_Lz_vs_logM}. 
Figure~\ref{f:Mcurr_vs_color} shows the resulting color-mass relations among
ancient GCs for the three IMFs. For comparison, we overplot the observed 
color-mass relations among metal-poor and metal-rich GC subpopulations in giant
elliptical galaxies taken from \citet[][i.e., their ``FCS+VCS high-mass''
galaxy sample]{mies+10}, after taking into account the dependence of $\cM/L_z$
on $\cM_{\rm GC}$ (and its dependence on the IMF) shown in
Figs.~\ref{f:MF_and_M_Lz_vs_logM}d\,--\,\ref{f:MF_and_M_Lz_vs_logM}f. 

Figures~\ref{f:Mcurr_vs_color}b and \ref{f:Mcurr_vs_color}c show that the 
observed color-mass relations among GCs in giant ellipticals are
significantly more pronounced than those predicted by the K09 GC disruption
models that use the Kroupa IMF (and, to a lesser extent, those using the
Salpeter IMF as well). This is especially true for the case of the metal-poor
GCs (i.e.,  the `blue tilt'). This is consistent with the findings of
\citet{mies+10} who used the older GC disruption models of \citet{krulam08} 
(which use the Kroupa IMF) to evaluate the impact of dynamical evolution to the
color-mass relations. Hence, \citeauthor{mies+10} concluded that dynamical
evolution of star clusters cannot be responsible for the shape of the blue
tilt seen in giant elliptical galaxies. However, the color-mass relations
predicted by the K09 models for $\alpha_{\it IMF} = -3.0$ are more similar to
the observed relations (see Fig.~\ref{f:Mcurr_vs_color}a). Specifically, the 
``average K09 model'' for $\alpha_{\it IMF} = -3.0$ reproduces the `red tilt'
observed among metal-rich GCs in giant ellipticals almost perfectly, while the
same model reproduces the `blue tilt' quite well for GC masses up to log
$\cM_{\rm GC} \simeq 6.1$, beyond which the observed blue tilt becomes more
pronounced than the K09 model prediction.

\section{The Nature of the GC Color-Mass Relation}  \label{s:CMR}

As shown in the previous section, dynamical evolution of GCs can
explain the shape of the color-mass relation among GCs in giant elliptical
galaxies up to log $\cM_{\rm GC} \simeq 6.1$ if the IMF was steep 
at subsolar masses ($-3.0 \la \alpha_{\it IMF} \la -2.4$), i.e., significantly
steeper than the Kroupa IMF for which $\alpha_{\it IMF} \simeq -1.7$ at those
masses. However, the fact that the blue tilt becomes more pronounced for log
$\cM_{\rm GC} \ga 6.1$ does not seem to be compatible with dynamical evolution
and requires a different explanation. 

In this context, previous studies already tested the potential efficacy of
stochastic fluctuations within the stellar CMDs, contamination of the GC
samples by tidally stripped dwarf galaxy nuclei, accretion of GCs found in
present-day low-mass galaxies, and field star capture by GCs
\citep{mies+06,miebau07}. None of these effects were found to be likely
contributors to the shape of the blue tilt in a significant manner.  

On the other hand, one effect known to have the potential to cause a
significant mass-metallicity relation among massive GCs is that of GC
self-enrichment by means of retention and reprocessing of SN\,{\sc ii} ejecta 
\citep[e.g.,][]{morlak89,recdan05,strsmi08,baihar09}. In the remainder of
this Section, we re-examine the GC self-enrichment scenario of
\citet[][hereafter BH09]{baihar09} and its expected effect on the mass
vs.\ \gz\ relation among GCs by taking into account the effects of
GC mass loss by both stellar evolution and dynamical evolution for
different IMF shapes. 

The latter effects can be divided up as follows: 
{\it (i)\/} the effect of GC self-enrichment on the \gz\ colors of GCs. This
  effect becomes increasingly important for GC masses above a certain minimum
  mass; 
{\it (ii)\/} the influence of GC mass loss on the present-day masses (and
  hence magnitudes) at which a given effect of GC self-enrichment is reached
  (including the dependence thereof on the adopted IMF); and 
{\it (iii)\/} the influence of dynamical evolution on the \gz\
colors of GCs (and its dependence on the adopted IMF). The latter effect
was shown and discussed above in Section \ref{s:gzevol}.

\subsection{The Impact of GC Mass Loss on GC Self-Enrichment} \label{s:selfenr}

We refer to BH09 for a full description of their GC self-enrichment model,
 including its assumptions, caveats and comparisons with other such
  models.  
Briefly, the BH09 model is built upon the assumptions that the energy
of the SN\,{\sc ii} ejecta is fully converted to kinetic energy, and that
self-enrichment occurs when this kinetic energy is equal to or less than the
binding energy of the primordial gas cloud that formed the GC. The 
metallicity of the protocluster cloud is parameterized as follows (see
Equation (7) of BH09): 
\begin{equation}
Z_c/Z_{\odot} = Z_{\rm pre}/Z_{\odot} + 10^{0.38+\log f_{\star} f_Z}
\label{eq:Z_C}
\end{equation}
where $Z_{\rm pre}$ represents the metallicity of the gas cloud prior to
self-enrichment, the factor 10$^{0.38}$ represents the SN yield used by BH09,
$f_{\star}$ represents the star formation efficiency (for which we adopt
BH09's value of 0.3), and $f_Z$ denotes the fraction of metals retained within
the cloud and incorporated during the formation of the lower-mass stars. In
deriving $f_Z$, BH09 define their ``reference model'' as one in which the
radial mass density profile of the protocluster cloud falls off as an
isothermal sphere (i.e., $\rho \propto r^{-2}$). For this case they find  
\begin{equation}
f_Z \approx \exp \left(-\frac{E_{SN} f_{\star} r_t}{10^{2} M_{\odot}\,G\,\cM_c} 
 \right){\rm ,} 
\label{eq:f_Z} 
\end{equation}
(see Equation 28 of BH09) where $E_{SN}$ is the energy released per SN\,{\sc ii}
(taken to be 10$^{51}$ erg), $r_t$ is the truncation radius of the
protocluster cloud (taken as 1 pc in BH09's reference model), and $\cM_C$
is the protocluster cloud mass. 

\begin{figure*}[tbhp]
\centerline{
\includegraphics[width=13.8cm]{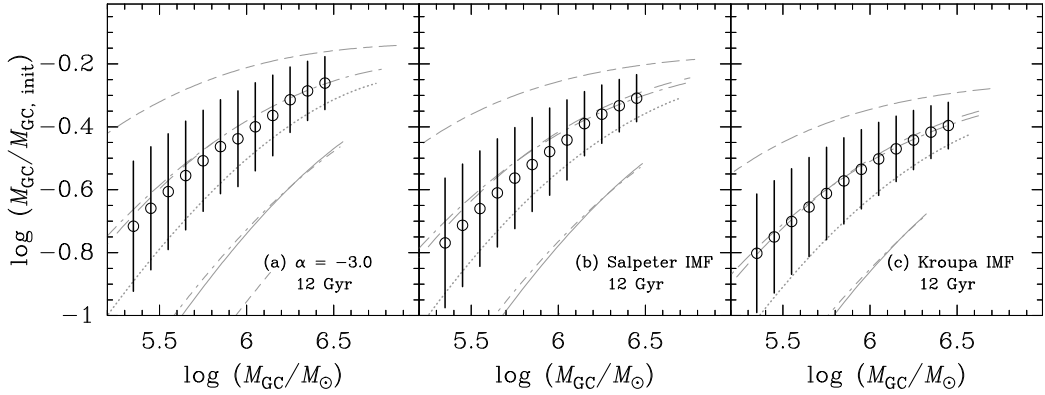}
}
\caption{Fractional remaining mass ($\cM_{\rm GC}/\cM_{\rm GC,\,i}$) versus
  current mass for GCs at an age of 12 Gyr. 
  {\it Panel (a)}: Thin lines of various styles represent the same K09
  models for $\alpha_{\it IMF} = -3.0$ as those shown in Fig.\
  \ref{f:MF_and_M_Lz_vs_logM}a with the same line styles.  
  Open circles represent the ``average'' models of K09 which were shown with
  thick black lines in Figure~\ref{f:MF_and_M_Lz_vs_logM}a.  
  {\it Panel (b)}: Similar to panel (a), but now for a \citet{salp55} IMF.
  {\it Panel (c)}: Similar to panel (a), but now for a \citet{krou01} IMF.
\label{f:Mcurr_Minit}
}
\end{figure*}

To estimate protocluster cloud masses from observed present-day GC masses, we
use  
\begin{equation}
\cM_{\rm GC} = \cM_{\rm GC, i} - \Delta \cM_{\rm GC}
\label{eq:M_GC_full}
\end{equation}
where $\cM_{\rm GC, i} = f_{\star} \, \cM_c$ is the initial GC mass and
$\Delta \cM_{\rm GC} \equiv \Delta \cM_{\rm pop} + \Delta \cM_{\rm dyn}$ is
the cumulative mass loss of the GC at an age of 12 Gyr, consisting of 
mass loss by stellar evolution ($\Delta \cM_{\rm pop}$) and by dynamical
evolution ($\Delta \cM_{\rm dyn}$). 
We determine values for $\Delta \cM_{\rm GC}$ by using the
``average'' models of K09\footnote{The K09 models use the
  \citet{mari+08} isochrones to evaluate $\Delta \cM_{\rm pop}$.} which
were shown by open circles in Fig.\ \ref{f:Mcurr_vs_color}. 
Figure~\ref{f:Mcurr_Minit} depicts $\cM_{\rm GC}/\cM_{\rm GC,\,i}$ 
versus $\cM_{\rm GC}$ for the three IMFs considered in this
paper. Note the significant dependence of the cumulative GC mass loss on the
IMF shape, especially for the most massive GCs. This effect is
mainly caused by $\Delta \cM_{\rm pop}$ being significantly larger for IMFs
with shallower slopes. 

Metallicities predicted by the BH09 reference model through equation
(\ref{eq:f_Z}) are converted to observed \gz\ colors by means of the
relation of \citet{peng+06a}, adopting $Z_{\odot}$ = 0.016
\citep{vand+07} and $\log\, Z/Z_{\odot}$ = [Fe/H] + 0.25 ($\pm$ 0.10) as found
for Galactic GCs as well as GCs in giant early-type galaxies 
\citep[e.g.,][]{shet+01,puzi+05,puzi+06,kirb+08}.  To compare model
predictions with the observed GC color-mass relations, we assume
pre-enrichment levels of [Z$_{\rm pre}$/H] = $-$1.45 and [Z$_{\rm pre}$/H] =
$-$0.25 for the metal-poor and metal-rich GC subpopulations,
respectively. 
These values yield \gz\ colors that agree with those of the observed red and 
blue tilts, respectively, at the low-mass limit of the GC sample.

\begin{figure*}[tbhp]
\centerline{
\includegraphics[width=14cm]{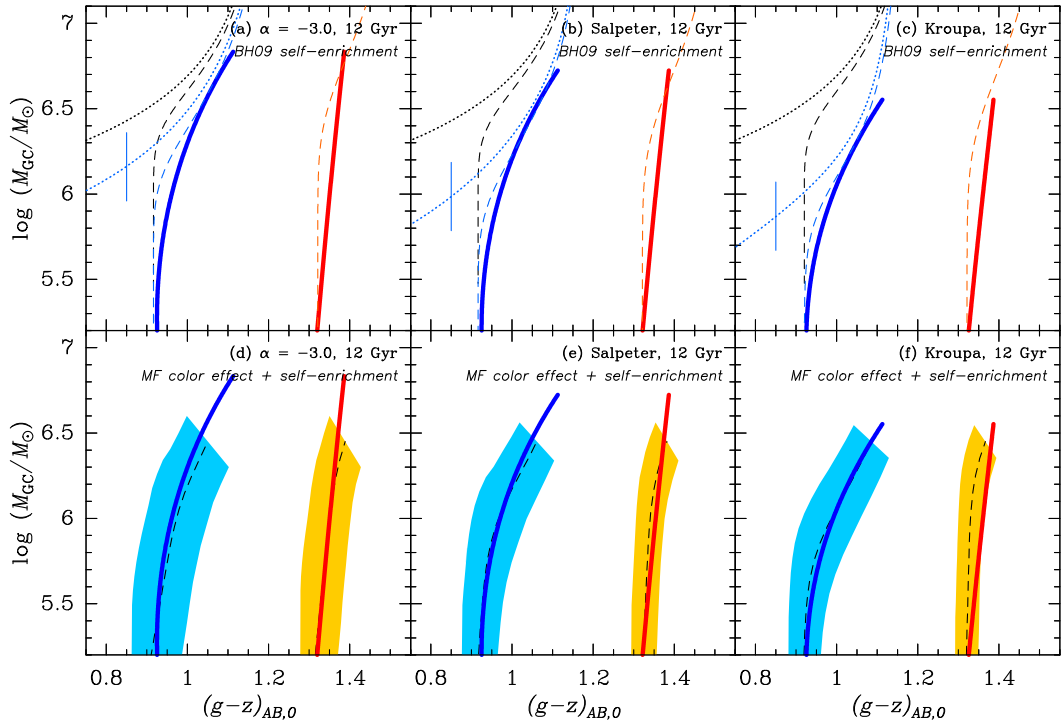}
}
\caption{Effects of self-enrichment and mass loss to GC color-mass 
  relations within high-mass early-type galaxies for the three
  IMFs considered in this paper.  
  {\it Panel (a)}: 
  Thick blue and red lines represent the observed color-mass relations
  among metal-poor and metal-rich GCs, respectively (cf.\
  Fig.~\ref{f:Mcurr_vs_color}). Thin black lines depict the reference
  GC enrichment model of BH09, while thin blue and orange lines do the
  same after taking into account 12 Gyr of GC mass loss according to
  the ``average'' K09 model for $\alpha_{\it IMF} = -3.0$. The blue
  vertical error bar on the left side of the panel indicates the
  uncertainty associated with the GC mass loss calculations (see
  Figure~\ref{f:Mcurr_Minit}). The dotted and dashed lines indicate the
  metallicity due to self-enrichment and the sum of self-enrichment and
  pre-enrichment, respectively, transformed into \gz\ color.  
  Blue and orange dashed lines assume pre-enrichment levels of
  log~$Z/Z_{\odot}$ = $-$1.45 and $-$0.25, respectively.   
  {\it Panel (b)}: Similar to panel (a), but now for a \citet{salp55} IMF.
  {\it Panel (c)}: Similar to panel (a), but now for a \citet{krou01} IMF.
  {\it Panel (d)}: Similar to panel (a), except that the dashed black
  lines now indicate the \gz\ colors caused by the \emph{cumulative} effects of GC
  mass-dependent $\alpha_{\rm MF}$ due to dynamical evolution 
  (depicted by the open circles in Fig.~\ref{f:Mcurr_vs_color}a) 
   \emph{and} GC pre- and self-enrichment (depicted by the blue and orange
   dashed lines in panel (a) above). Shaded 1-$\sigma$ contours
     around the dashed black lines depict the uncertainties 
   associated with the [Z/H]\,--\,color transformation and the GC mass loss
   calculations.  
  {\it Panel (e)}: Similar to panel (d), but now for a \citet{salp55} IMF.
  {\it Panel (f)}: Similar to panel (d), but now for a \citet{krou01} IMF.
  See Sections~\ref{s:selfenr} and \ref{s:combined} for details. 
\label{f:massloss_all}
}
\end{figure*}

In the top panels of Figure~\ref{f:massloss_all}  (i.e.,
Figs.~\ref{f:massloss_all}a\,--\,\ref{f:massloss_all}c) we plot the color-mass 
relations that would be expected if GCs are self-enriched according to the
reference BH09 model, after which the GCs undergo 12 Gyr of mass loss
according to the ``average'' K09 models for the three different IMFs. 
These diagrams are based on Fig.~\ref{f:Mcurr_vs_color} after removing the 
 symbols associated with the K09 model predictions. 
The dotted lines indicate the color-mass relation due purely to
self-enrichment, while the dashed lines 
do so for the case of pre-enrichment \emph{plus}
  self-enrichment. 
Each set of lines is drawn twice (using different pen colors) to indicate the
predictions with and without the effect of GC mass loss.  

Concentrating on the results for the metal-poor GC subpopulation, we see that
the effect of GC mass loss on the BH09 model predictions is significant. 
Specifically, the color-mass relation predicted by the standard BH09 model is
consistent (to within the uncertainties) with the observed color-mass relation
from \citet{mies+10} for Kroupa and Salpeter IMFs. For the $\alpha_{\it IMF} =
-3.0$ IMF, the predicted colors are somewhat bluer than the observed
color-mass relation in the mass range $5.6 \la \log\,(\cM_{\rm GC}/M_{\odot}) \la
6.3$. 
Conversely, looking at the predictions for the metal-rich GC subpopulation, we
see that the effect of self-enrichment is expected to become observable only
for current GC masses $\log\,(\cM_{\rm GC}/\Modot) \ga 6.4$, whereas the red
tilt is observed down to significantly lower GC masses.  

\subsection{Combining the Effects of Self-Enrichment and Dynamical
  Evolution} \label{s:combined} 
The combined effects of GC self-enrichment and the evolution of the stellar
mass function in GCs on their \gz\ colors are illustrated in the bottom panels
of Figure~\ref{f:massloss_all} (i.e.,
Figs.~\ref{f:massloss_all}d\,--\,\ref{f:massloss_all}f). 
This is done by first converting the GC mass-dependent \gz\ colors due to 
the effect of dynamical evolution of GCs on $\alpha_{\rm MF}$ (shown in
Figs.~\ref{f:MF_and_M_Lz_vs_logM}a\,--\,c and \ref{f:color_vs_alpha}) to
metallicity space using the transformation of
\citet{peng+06a}. To this we add the metallicity due to GC self-enrichment
(i.e., the equivalent of the blue dotted lines in
Figs.~\ref{f:massloss_all}a\,--\,c). The total metallicities are then
converted back to \gz\ using the \citet{peng+06a} transformation as before.  

Figs.~\ref{f:massloss_all}d\,--\,\ref{f:massloss_all}f reveal an interesting
result for the metal-poor GC subpopulation: when adding the simulated effects
of GC self-enrichment (including 12 Gyr of GC mass loss) and the evolution of
$\alpha_{\rm MF}$ in GCs on their \gz\ colors, the resulting color-mass
relation becomes consistent with the observed blue tilt among metal-poor GCs
in giant ellipticals {\it for all IMF shapes considered here}. In fact,
differences between results that use different IMF shapes are smaller than the
uncertainties in this respect. This result is due to the fact that
while the effect of the mass-dependent evolution of $\alpha_{\rm MF}$ (due to
dynamical evolution of GCs) on the \gz\ colors is significantly stronger for
steep IMFs (e.g., $\alpha_{\rm IMF} = -$3.0) than for Salpeter or Kroupa IMFs,
the opposite is true for the effect of GC mass loss due to stellar evolution
(and hence the impact of GC self-enrichment for GCs with $\cM_{\rm GC} \ga
10^{6.1}\;\Modot$), and the two effects cancel out each other to within the
uncertainties. 

The situation is different for the metal-rich GC subpopulation. The 
effect of GC self-enrichment on the \gz\ colors of metal-rich GCs is much
weaker than for metal-poor GCs: our calculations indicate that only
the most massive metal-rich GCs (with $\cM_{\rm GC} \ga 10^{6.4}\;\Modot$)
will have had their \gz\ colors slightly affected by self-enrichment (by
$\Delta\,g\!-\!z \la 0.05$ relative to less massive GCs). 
Conversely, the effect of the mass-dependent evolution of $\alpha_{\rm MF}$
on the \gz\ color causes an approximately linear color-mass relation over the
full range of GC masses considered, and the tilt of this relation is only
significant for bottom-heavy IMFs (cf.\ Section~\ref{s:gzevol}).  
The presence of a color-mass relation among metal-rich GCs in giant
ellipticals down to masses $\cM_{\rm GC} \la 10^{6}\;\Modot$, as seen in the
large sample of \citet{mies+10}, therefore seems to indicate that the
metal-rich GCs in giant ellipticals were formed with a
bottom-heavy IMF ($\alpha_{\rm IMF} \la -2.3$). This result reinforces 
the conclusions of Paper I as well as those of various recent galaxy studies
based on gravity-sensitive spectral features in the red part of the optical
spectrum \citep[e.g.,][]{vancon10,convan12b,smit+12} 
and dynamical $\cM/L$ values that exceed those predicted by stellar
population models based on a Kroupa IMF \citep[e.g.,][]{capp+12,tort+13,lask+13}. 

\subsection{Comparison to Previous Studies} \label{s:comp}

As to the calculations of the effects of GC self-enrichment done in Section
\ref{s:selfenr}, the main differences between our results and those of 
both BH09 itself and \citet{mies+10} are related to the treatment of GC mass
loss. Specifically, BH09 did not take GC mass loss into account when comparing
their simulations with observed GC color-magnitude relations in giant
elliptical galaxies. \citet{mies+10} did implement an aspect of GC mass loss,
using the GC dissolution scenario of \citet{jord+07b} to estimate GC mass
loss. However, the latter study only considered GC mass loss due to two-body
relaxation and hence did not incorporate the important effect of mass loss due
to stellar evolution ($\approx$\,40\% of the initial mass for a Kroupa or
Chabrier IMF at an age of 12 Gyr: see, e.g., \citealt{bc03} or
\citealt{mara05}; see also Fig.~\ref{f:Mcurr_Minit}). The impact of a full
treatment of GC mass loss (and its dimming effect) on self-enrichment
estimates is significant. 
One illustration of this is that \citet{mies+10} noted two main apparent problems 
with their implementation of the reference model of BH09 in reproducing the
blue tilt seen among GCs in giant ellipticals: {\it (i)\/} the model predicts 
self-enrichment to become important only at masses $\ga 2 \times 10^6\;
\Modot$ whereas the blue tilt starts at a mass that is an order of magnitude
lower, and {\it (ii)\/} the mass-metallicity relation predicted by the
model is significantly more pronounced in the self-enrichment regime 
than the observed blue tilt. 
Consequently, \citet{mies+10} looked into changes to the assumptions of
the radial density profile and star formation efficiency in the reference BH09 
model to improve the agreement of the model predictions with the data. 
In particular, Mieske et al.\ needed to make the radial density profile of the
protocluster cloud significantly steeper than an isothermal profile to make
the self-enrichment model fit the observed MMR, which seems physically
unlikely.
However, as shown in Fig.~\ref{f:massloss_all}, the implementation of a full
treatment of GC mass loss (for which the impact to the mass-metallicity
relation depends on GC mass as well as the IMF) renders a situation where the 
reference BH09 model reproduces the blue tilt exhibited by the massive GCs
very well. 

\subsection{Dependence on Model Assumptions and Ingredients}

In this Section we assess to what extent our results shown in
Fig.~\ref{f:massloss_all} might depend on the assumptions and ingredients of
the color-metallicity relation and the GC dynamical evolution model used to derive them.  

\subsubsection{The Color-Metallicity Relation}

\begin{figure*}[tbhp]
\centerline{
\includegraphics[width=14cm]{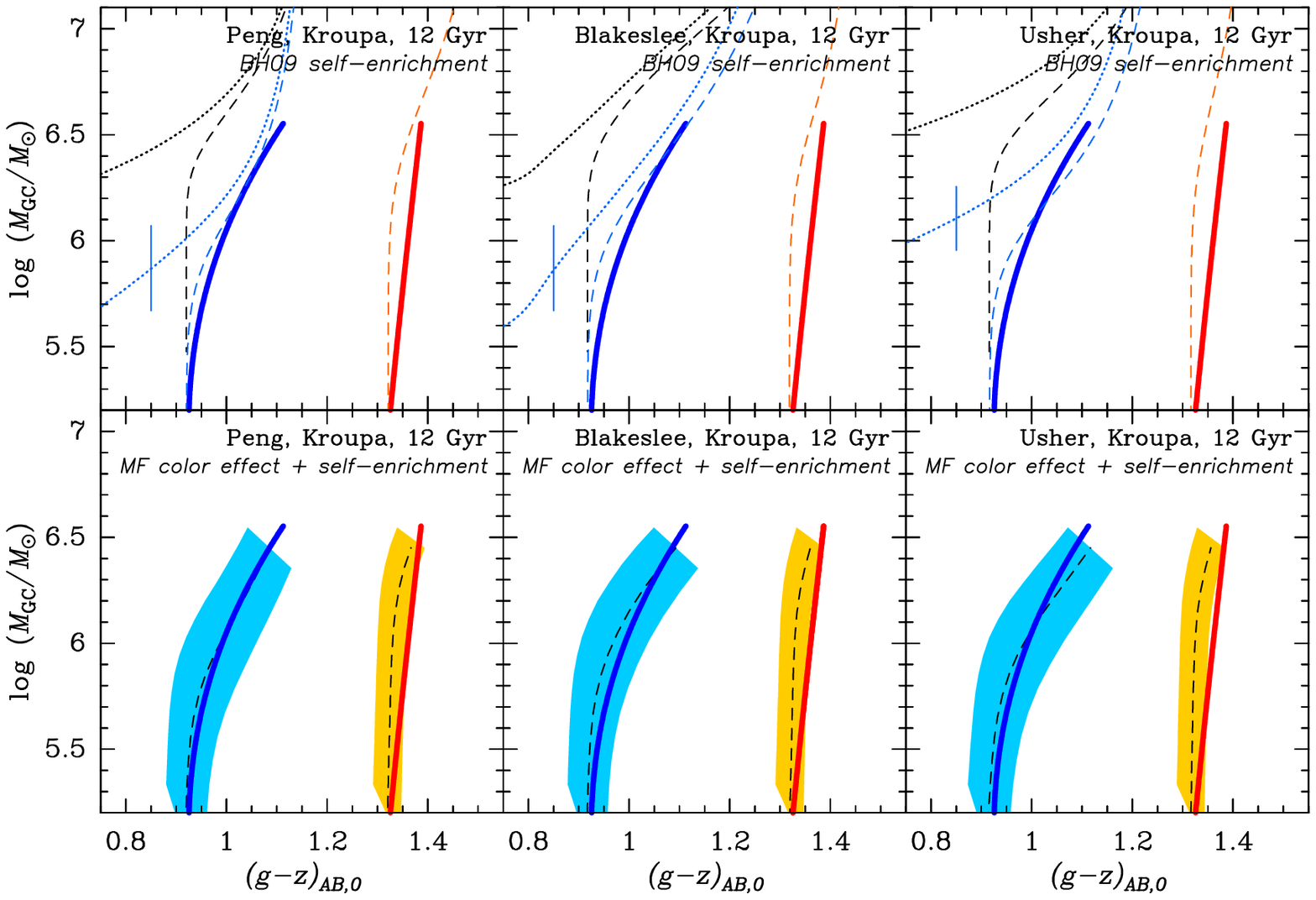}
}
\caption{Illustration of the impact of different color-metallicity relations
  to the color-mass relations. The three pairs of top \& bottom panels are
  copies of panels (c) and (f) of Figure~\ref{f:massloss_all} using the
  color-metallicity relations of \citet[][left panels]{peng+06a}, 
  \citet[][middle panels]{blak+10}, and \citet[][right panels]{ushe+12}. Note
  that the predicted color-mass relations among the metal-poor 
  and metal-rich GC subpopulations are largely insensitive to the choice of
  color-metallicity relation. 
\label{f:CMcompare1}
}
\end{figure*}

To translate metallicity variations into \gz\ color changes in Section
\ref{s:selfenr}, we followed \citet{mies+10} and used the broken linear
function of \citet{peng+06a} which was derived from integrated-light $g$ and
$z$-band photometry of 95 GCs in the Milky Way, M49, and M87 that have
spectroscopic metallicities. 
However, the scatter in the Peng et al.\  relation is significant, especially
at intermediate metallicities  ($-1.2 \la {\rm [Z/H]} \la -0.5$) and
supersolar ones, and uncertainties of several individual data points are
substantial. To assess the uncertainty associated with the choice of a
particular [Z/H]\,--\,(\gz) relation, we apply two other such relations from the
recent literature for comparison purposes: 
\emph{(i)} the quartic polynomial function of \citet{blak+10}, which was
derived from the same data as the \citet{peng+06a} function, and 
\emph{(ii)} the broken linear function of \citet{ushe+12}, which was derived
from an independent empirical database of colors and spectroscopic
metallicities of 903 extragalactic GCs. 

As might be expected, the three [Z/H]\,--\,(\gz) relations imply different
amounts of GC pre-enrichment needed to let the mean color of the modeled blue
and red tilts agree with the observed ones. Specifically, the \citet{blak+10}
relation implies [Z$_{\rm pre}$/H] = $-$1.37 and [Z$_{\rm pre}$/H] = -0.22 for
the blue and red GCs, respectively, while the [Z$_{\rm pre}$/H] values are
$-$1.20 and $-$0.18 when using the \citet{ushe+12} relation. 

The sensitivity of our results to the choice of a given different
color-metallicity relation is illustrated in Figure~\ref{f:CMcompare1}, which
consists of three copies of Figs.\ \ref{f:massloss_all}c and
\ref{f:massloss_all}f, one for each color-metallicity relation.  
While the resulting shapes of the blue and red
tilts depend somewhat on the color-metallicity relation in a quantitative
sense, they are all consistent with one another to within the 1-$\sigma$
uncertainties. Moreover, our main results do not change: The addition of the
effect of 12 Gyr of GC mass loss to that of self-enrichment yields a good fit
to the observed blue tilt, while the shape of the observed red tilt indicates
a bottom-heavy IMF at sub-solar masses for the metal-rich population.  

\subsubsection{The GC Dynamical Evolution Model}

\begin{figure}[tbhp]
\centerline{
\includegraphics[width=6.7cm]{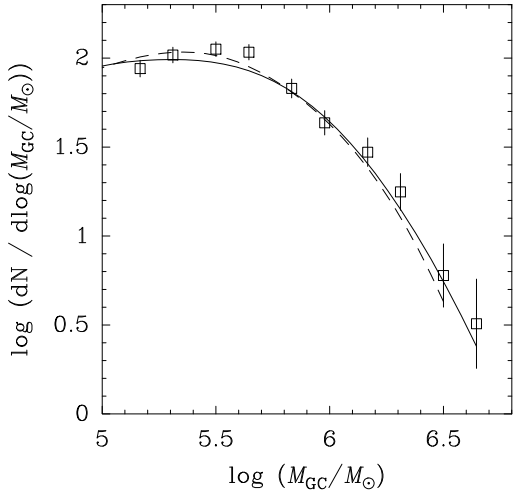}
}
\caption{GCMF of the Galactic GC system \citep{harr96} for $\log\,\cM_{\rm GC}
  \geq 5$, assuming $\cM/L_V = 3$ (open squares and 1-$\sigma$ error
  bars). The solid line represents the GCMF calculated from the GC luminosity
  function derived by \citet{krupor09} using a model involving
  mass-dependent $\cM/L$ ratios and dissolution timescales $t_0$ determined
  for each individual GC (see their Fig.\ 3). The dashed line represents a fit
  to the Galactic GCMF using a model assuming a mass-independent $\cM/L$ ratio
  \citep[as in][]{falzha01}. Data are from \citet{krupor09}. Note the
  similarity of the two curves in the mass range of GCs considered in this
  paper.  
\label{f:MFcompare}
}
\end{figure}

To estimate the variation in the effects of 12 Gyr of GC mass loss among
different GC dynamical evolution models, we compare the K09 model with that 
of \citet[][hereafter FZ01]{falzha01}. The main difference between
these two models is that FZ01 employs a constant, GC mass-independent value
for the (present-time) $\cM/L$ to compare model masses with observed
luminosities, while K09 uses GC mass-dependent mass loss rates that result in
GC mass-dependent $\cM/L$ rates (see Fig.~\ref{f:MF_and_M_Lz_vs_logM}). The
two models also use different parametrizations of GC mass loss by two-body
relaxation, and K09 incorporates a prescription of the effect of stellar
remnant retention which FZ01 did not. However, in spite of these different
model ingredients, the only significant differences between the GC mass
functions (GCMFs) for the Galactic GC system at an age of 12 Gyr as inferred
by the two models show up at \emph{low} GC masses \citep[$\cM_{\rm GC} \la 3
\times 10^4 \; \Modot$, see][]{krupor09}.  
For the massive GCs considered in this paper ($\cM_{\rm GC} \ga 2
\times 10^5 \; \Modot$), the difference between the GCMFs inferred by the two
models are only of order 10\% for any given mass bin, which is within the
1\,$\sigma$ Poisson error bars (see Figure \ref{f:MFcompare}). 
Since these differences are smaller than the differences
between K09 models with different values of dissolution timescales $t_0$
employed in this paper (illustrated by the blue error bars in
Figs.~\ref{f:massloss_all}a\,--\,\ref{f:massloss_all}c), we believe our results
on the amounts of GC mass loss, and their uncertainties, are robust. 

\subsection{Effect of GC Dissolution Timescale} \label{s:t_0}

As mentioned in the introduction, the color-mass relations among GCs show 
environmental dependencies \citep{mies+10}. Firstly, GCs associated with
high-mass galaxies exhibit more pronounced blue \emph{and} red tilts than do GCs in
lower-mass galaxies. Secondly, the blue tilt is found to be more pronounced
among GCs at smaller galactocentric distances than among those further out. 
One might expect such environmental dependences to exist among ancient GC
systems due to dynamical evolution (i.e., tidal shocks and two-body
relaxation), as tidal fields are stronger in more massive galaxies and at
smaller galactocentric distances. A recent illustration of this effect was
shown by \citet{madr+12} who performed N-body simulations of GCs in a realistic
Milky Way-like potential and found a strong trend of increasing mass loss with
decreasing galactocentric distance for a given type of GC
\citep[initial mass, half-mass radius, concentration; see also][]{vesheg97}. 

\begin{figure*}[tbhp]
\centerline{
\includegraphics[width=14cm]{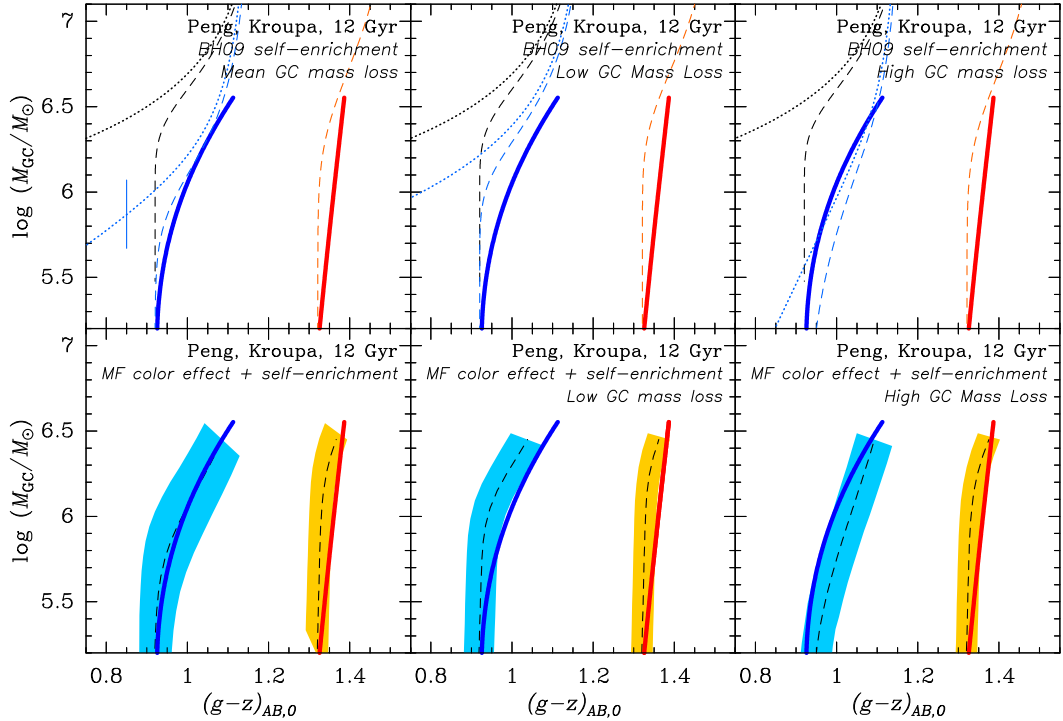}
}
\caption{Illustration of the impact of different amounts of GC mass loss 
  to the color-mass relations. The three pairs of top \& bottom panels are copies of panels
  (c) and (f) of Figure~\ref{f:massloss_all} using three percentiles within
  the range of GC mass loss due to 12 Gyr of dynamical evolution found in our modeling:
  50\% (i.e., as in Fig.\ \ref{f:massloss_all}; left panels), 20\% (middle
  panels), and 80\% (right panels). Note that the shape of the blue tilt is
  sensitive to the cumulative amount of GC mass loss due to dynamical evolution, while
  the red tilt is much less so.  
\label{f:CMcompare2}
}
\end{figure*}

To simulate the effect of environmentally varying GC dissolution timescales on
the color-mass relations among GCs for a given IMF, we select GCs whose
cumulative mass loss over 12 Gyr of stellar and dynamical evolution are at the 20\% and 
80\% percentiles of the range found among the GC families considered in this
paper (see Figure~\ref{f:Mcurr_Minit}c), and compare the resulting color-mass
relations with that derived for the ``average'' K09 model. This comparison is
depicted in Figure~\ref{f:CMcompare2} for the case of the Kroupa IMF and the
\citet{peng+06a} color-metallicity relation. It is obvious that the shape of
the blue tilt is indeed sensitive to the amount of cumulative GC mass loss
(and its impact on self-enrichment in the observed color-mass frame):
the blue tilt only shows up for $\log \cM_{\rm GC} \ga 6.1$ for the ``low''
mass loss case while it is present and pronounced all the way down to the
minimum mass of the GC sample for the ``high'' mass loss case. 
Note that this behavior is entirely consistent with the observed environmental
dependencies of the blue tilt found by \citet{mies+10}. 

Note also that in contrast with the blue tilt, the dependence of
the shape of the \emph{red} tilt on the cumulative amount of GC mass loss is
negligible in terms of the impact of GC mass loss on self-enrichment in the
observed color-mass frame. This suggests 
that the presence of the red tilt in massive ellipticals and its absence in
low-mass ellipticals is \emph{not} mainly due to differences in cumulative GC
mass loss. Instead, we suggest that it indicates that metal-rich GCs in
massive ellipticals were formed with a bottom-heavy IMF, as already mentioned
in Section~\ref{s:combined}.

\subsection{Differences in Blue Tilts Between Giant Elliptical Galaxies}

Several studies of the blue tilt in giant elliptical galaxies have pointed out
that the color-magnitude relations among metal-poor GCs can vary significantly
from galaxy to galaxy, even among massive galaxies with similar luminosities. A
well-known case in point is M49, the brightest giant elliptical galaxy in the
Virgo cluster, which does not show an obvious blue tilt whereas NGC 1399 and
M87, two other giant elliptical galaxies with similar luminosities, do
\citep{stra+06,mies+06,mies+10}. 
While such differences might seem difficult to reconcile within a scenario in
which the mass-metallicity relation among massive GCs is primarily due to
self-enrichment, one should consider the following. 

First and foremost, we emphasize that the main purpose of this study has been
to compare calculations of the impact of GC mass loss and self-enrichment to
the {\it average\/} color-mass relations among GCs in a large sample of giant
elliptical galaxies. Several factors render averaging over many galaxies a
necessity in this context.  As shown and discussed above, $\sim$\,12 Gyr of
dynamical evolution can cause several differences between different families
of GCs (especially for the case of bottom-heavy IMFs). 
One example of such differences is the relative amount of mass lost
(see Section~\ref{s:t_0} and Figs.~\ref{f:Mcurr_Minit} and
  \ref{f:CMcompare2}).  
This results in a spread of current luminosities among GCs that started out
with a given initial mass, and hence with a similar amount of
self-enrichment. Another property that is quite sensitive to the initial GC
conditions is the change of the slope of stellar mass function at subsolar
masses and its associated change of integrated \gz\ color after 12 Gyr (see
Figs.~\ref{f:MF_and_M_Lz_vs_logM}a\,--\,c and \ref{f:color_vs_alpha} as well
as Paper I). The 1-$\sigma$ uncertainties associated with 
these two effects in the color-mass plane are indicated in
Figs.~\ref{f:massloss_all}d\,--\,f. Last but not least, there is a significant  
scatter among colors of metal-poor GCs in individual giant ellipticals
($\sigma\,{\rm [Z/H]} \simeq 0.3$; e.g., \citealt{harr+06}), which translates
to $\sigma (g\!-\!z) \simeq 0.15$. This is most likely mainly due to a spread
in pre-enrichment levels (e.g., BH09). For GCs near the massive end of the GC mass
function (where GCs are scarce and hence stochastic variations between GCs are
significant), the impact of this scatter on the resulting color-mass relation
can be significant, especially for galaxies with relatively small specific GC
frequencies. Note in this respect that M49 hosts significantly fewer GCs
  than NGC 1399 and M87 do, see Figure 3 of \citet{mies+10}. 

That said, important steps forward in the context of understanding the nature
of the color-mass relations of GCs as well as their variations between
galaxies can already be made with currently available instrumentation. One
such step will be to obtain accurate photometry and size measurements of GCs  
{\it outside the central few kpc\/} sampled by the currently
available HST observations of nearby giant ellipticals (typically only a
central pointing). 
This will not only yield color-magnitude data for GCs in
weaker tidal fields which will allow one to evaluate differences in sizes and
hence in dynamical evolution \citep[e.g.,][]{falzha01,mclfal08,goud12}. The lower
background levels of the underlying diffuse galaxy light in the outer regions
will also allow high-S/N measurements of spectroscopic metallicities of GCs in
the upper $\sim$\,3 mag of the GC luminosity function with 8-10\,m class
telescopes \citep[e.g.,][]{puzi+05,cena+07}. The latter will be important to
decipher the true color-metallicity relation 
(rather than one derived from colors and [Fe/H] values of Galactic GCs; for
example, the few Galactic GCs that are moderately metal-rich ($[Z/H] \ga -0.5$)
have significant foreground extinction) and its impact on the nature of the blue tilt.

\section{Summary and Conclusions} \label{s:conc}

Driven by several lines of recent evidence for a ``bottom-heavy'' stellar
IMF in massive elliptical galaxies, we have investigated the influence of the
shape of the IMF on the color-magnitude relations for the metal-poor and
metal-rich subpopulations of GCs in such galaxies. To this end we used mass
loss calculations for GCs for three IMF shapes and a large range of initial GC
masses, in conjunction with calculations of integrated-light colors from
isochrones.  These calculations were used for two main purposes: 
{\it (i)\/} to evaluate the effect of evolution of the stellar
mass function at subsolar masses on the color-magnitude relations, and 
{\it (ii)\/} to evaluate the impact of GC mass loss on the
\citet[][BH09]{baihar09} model of GC self-enrichment when comparing model
predictions to observed color-magnitude diagrams. Our results are compared to
observed color-magnitude relations among a very large set of GCs in giant
elliptical galaxies from \citet{mies+10}. Our main conclusions are as follows. 
\begin{itemize}
\item 
As to the observed color-magnitude relation among the metal-rich GCs in giant
elliptical galaxies (often referred to as the ``red tilt''), we found that
the effect of dynamical evolution of GCs on the stellar mass functions over a
Hubble time can explain the observed slope of the red tilt along its full
extent {\it if (and only if) the IMF was bottom-heavy\/} 
(e.g., $-3.0 \la \alpha \la -2.3$ in $dN/d\cM \propto \cM^{\alpha}$ at
subsolar masses). \\ [-3.3ex]
\item 
Conversely, we find that GC self-enrichment is only able to cause a small 
reddening for the most massive metal-rich GCs (i.e., current masses $\cM_{\rm GC}
\ga 3 \times 10^6\; \Modot$). Hence we argue that the observed color-magnitude
relation among metal-rich GCs in massive elliptical galaxies constitutes
evidence for a bottom-heavy IMF in such galaxies. \\ [-3.3ex]
\item 
As to the ``blue tilt'' among metal-poor GCs, we find that the influence of
dynamical evolution of GCs on the stellar mass functions over a 
Hubble time can reproduce the observed tilt only up to a current GC mass 
$\cM_{\rm GC} \simeq 2 \times 10^6 \; M_{\odot}$, and only if the IMF was
bottom-heavy. Beyond this mass, the observed blue tilt is more pronounced than
what can be explained by the evolution of stellar mass functions in
GCs, so that additional effects (such as self-enrichment) are still required
to reproduce the full extent of the observed blue tilt.  \\ [-3.3ex]
\item 
When considering the dimming effect of 12 Gyr of mass loss in color-magnitude
diagrams of GCs, we find that the ``reference'' GC self-enrichment model of
BH09 reproduces the shape of the observed blue tilt very well in the cases of 
Kroupa or Salpeter IMFs. This result differs from those of previous studies,
which did not take GC mass loss by stellar evolution into account.  \\ [-3.3ex]
\item
When adding the simulated effects of GC self-enrichment and the evolution of
stellar mass functions in metal-poor GCs on their \gz\ colors, the resulting
color-mass relation becomes consistent with the observed blue tilt {\it for
  all IMF shapes considered here}. This result is due to the fact that while
the effect of the GC mass-dependent evolution of stellar mass functions to the
\gz\ colors is significantly stronger for steep IMFs (e.g., $\alpha 
= -$3.0) than for Salpeter or Kroupa IMFs, the opposite is true for the effect
of GC mass loss due to stellar evolution, and the two effects cancel out each
other to within the uncertainties. \\ [-3.3ex]
\item 
 We looked into the sensitivity of our results to the choice of the 
  color-metallicity relation. While details
  such as the mean pre-enrichment metallicity required to make our modeled
  color-mass relations fit the observed blue and red tilts do differ slightly
  among color-metallicity relations from the recent literature, the
  \emph{shapes} of the modeled color-mass relations are found to be consistent
  from one color-metallicity relation to another. Similarly, the sensitivity
  of our results to the choice of GC dynamical evolution model was found to be
  insignificant in the relevant range of GC masses. \\ [-3.3ex]
\item
 Differences in cumulative GC mass loss expected for different galactic
  environments (e.g., inner versus outer regions of massive galaxies) are
  found to affect the shape of the blue tilt in a way that is qualitatively
  consistent with the observed variations. However, the impact of varying GC
  mass loss on the shape of the \emph{red} tilt is found to be negligible. 
\end{itemize}

We argue that further insights into the color-magnitude relations of GCs as
well as their variations between galaxies can be obtained by means of
high-resolution imaging as well as high-S/N spectroscopy of GCs outside the
central few kpc sampled by the currently available HST observations of nearby
giant ellipticals.  
With such information, one will be able to constrain the true color-metallicity
relation as well as environmental influences on the properties of the blue tilt.  

\acknowledgments
We acknowledge useful discussions with Jeremy Bailin, Charlie Conroy, Mike
Fall, Mark Krumholz, and Enrico Vesperini.  
We highly appreciated the referee's insightful and constructive 
comments which improved the paper.
We made heavy use of the SAO/NASA Astrophysics Data System while writing this paper. 
PG was partially supported during this project by NASA through grant
HST-GO-11691 from the Space Telescope Science Institute, which is operated by
the Association of Universities for Research in Astronomy, Inc., under
NASA contract NAS5--26555. JMDK acknowledges the hospitality of the Aspen
Center for Physics, which is supported by the National Science Foundation
through Grant No.\ PHY-1066293.



\end{document}